\documentclass{amsart}
\usepackage{graphicx}
\usepackage{amsmath}

\newtheorem{theorem}{Theorem}

\newtheorem{lemma}[theorem]{Lemma}

\newtheorem{proposition}[theorem]{Proposition}

\begin{document}
\title{The Measurement of Time}
\thanks{The research of the authors was supported by NSERC grants. }
\author{Abraham Boyarsky}
\address{Department of Mathematics and Statistics, Concordia University, 1455 de
Maisonneuve Blvd. West, Montreal, Quebec H3G 1M8, Canada}
\email{boyar@alcor.concordia.ca}
\author{Pawe\l \ G\'ora}
\address{Department of Mathematics and Statistics, Concordia University, 1455 de
Maisonneuve Blvd. West, Montreal, Quebec H3G 1M8, Canada}
\email{pgora@mathstat.concordia.ca}
\date{\today }

\begin{abstract}
We present a definition of time measurement based on high energy photons and
the fundamental length scale and show that, for macroscopic time, it is in
accord with the Lorentz transformation of special relativity. To do this we
define observer in a different way than in special relativity.
\end{abstract}

\maketitle

\pagestyle{myheadings} \markboth{ABRAHAM BOYARSKY and  PAWE\L \ G\'ORA 
}{DEFINITION OF MEASURABLE TIME}

\section{ Introduction:}

String theory and loop quantum gravity theory claim that space and time are
ultimately discrete. In spite of this, however, there has not been a serious
attempt to derive the continuum equations of General and Special Relativity
from a discrete space perspective. The main objective of this note is to
present a definition of time measurement based on the fundamental Planck
length scale that, on macroscopic scales, is consistent with the Lorentz
transformation that characterizes Special Relativity. To carry this out we
define time as a property of space rather than as an independent coordinate.
In our definition, time is measured at a spatial location of Planck length
by the amount of energy of very high frequency photons it receives.

The notion of observer in Special Relativity requires synchronized clocks 
throughout space and an experimenter
who has complete and immediate access to all the information of the
synchronized clocks \cite[p. 78]{4.}. In this note we present a different
 method of
time observation.

\section{ Model for Time Measurement}

We assume that space has a fundamental length scale, $L_{P},$ the Planck
length. Furthermore, we assume that the fundamental length scale is observer
independent as in doubly-special relativity, DSR, [1-3]. That is, any moving
frame measures $L_{P}$ as the smallest unit of length. If this were not the
case then length contraction would contravene the notion of minimum length.
With velocity limited by the speed of light, we have a minimum time 
unit, $T_{p}.$ Time
dilation can only increase $T_{p}$. Hence no assumption is necessary on a
frame independent fundamental scale for time. We now state
\medskip

Postulate 1: The Planck length is the smallest unit of length and this
length is frame independent.
\medskip

Einstein assumed that a photon is a bundle of energy localized in a small
volume of space, and that it remains localized as it moves away from the
source with velocity $c$. The Planck-Einstein equation relates the energy
content $E$ of the photon to its frequency $\nu $ by the equation 
\ $E=h\nu.$ But what is meant by the frequency of a photon if it is 
localized in space? Since $\nu =c/\lambda $, where $\lambda $ is the
wavelength, a photon of frequency $\nu $ is one whose entire energy
 content is contained within a length $\lambda =c/\nu ,$ and
the illusion of a wave is created by a stream of photons with the same
frequency$.$ From now on we let $\lambda =L_{P}$ and refer to such 
photons
as photons of maximal energy or simply maximal photons. Then $\nu
_{P}=c/L_{P}$ is the frequency corresponding to the smallest wavelength
possible, and possessing the maximum energy possible: 
$E_{P}=hc/L_{P}=h/T_{p}= 7.68 10^28 eV.$ 
Particles with very high energy (UHE) are known to exist [5-7].

In summary, we view a photon of wavelength $L_{P}$ to mean that all of its
energy is localized in $L_{P}.$ It is convenient this think of this energy
as the total area under a power form such as a normal density supported
entirely inside a Planck length as depicted in Figure 1. The time it takes
for this power form - moving at the speed of light - to pass through $L_{P}$
is $T_{P}.$ We summarize the foregoing in

\medskip

Postulate 2: The maximum energy a photon is $E_{P}=hc/L_{P}.$ At any
location having Planck length, this energy allows the measurement of an
amount of time equal to $T_{P}=h/E_{P}=L_{P}/c.$
\medskip

We regard the time duration of an event as a property of the spatial
location from where the event was observed. Let us consider the location
 $[0,L_{P}]$ in a stationary frame S as depicted 
in Figure 2. An event spanning $N$ energy pulses is shown in Figure 2,
 resulting in $NT_{P}$ units of time being measured at the location 
$[0,L_{P}]$. The energy graph shown in Figure 2 is the sum of the
 energy graphs of $N$ maximal photons.
\medskip
Postulate 3: The time duration of an event measured at a location of Planck
length in any frame is the number of energy pulses $E_{P}$ it receives
multiplied by $T_{P}$.

\medskip

\begin{center}
\centerline{\includegraphics[bb=101 553 425 725,width=4.67in,
height=2.0in,keepaspectratio]{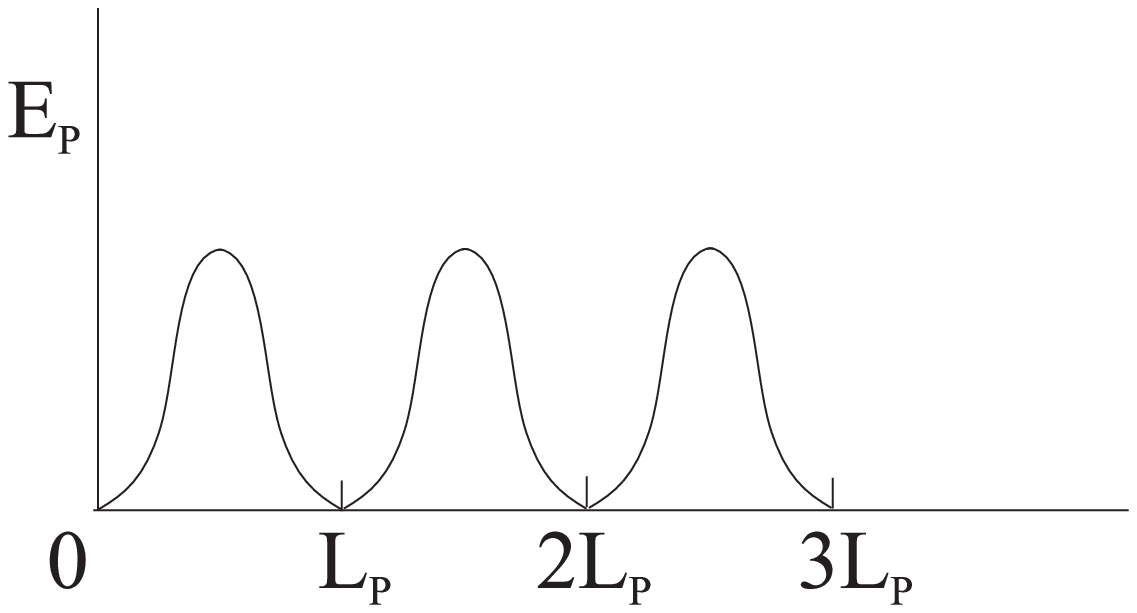}} \bigskip Figure 1: A maximal photon
wave 
\end{center}

\bigskip

\begin{center}
\centerline{\includegraphics[bb=83 553 286 725,width=2.02in,
height=2.02in,keepaspectratio]{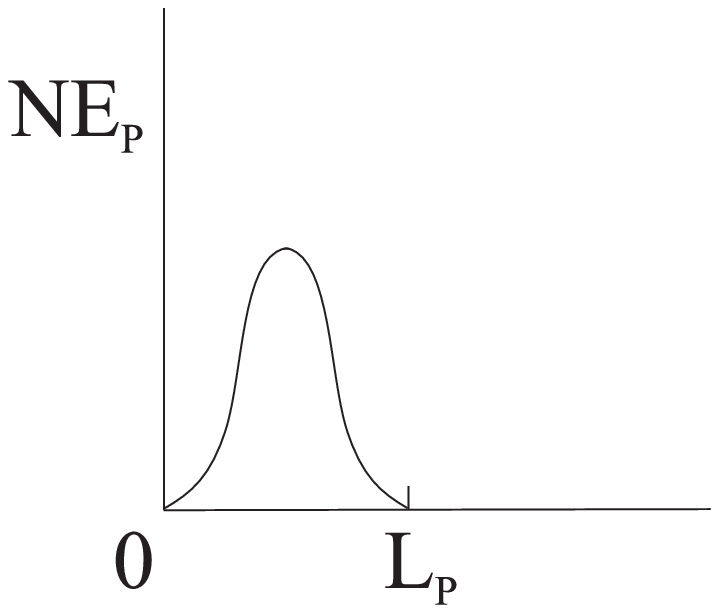}}

\bigskip Figure 2: An event of duration $NT_{P}$ at location $[0,L_{P}]$
interpreted as the reception of $N$ pulses of energy from a stream of
maximal photons.

\bigskip
\end{center}

\section{ Special Relativity}

We now show that the foregoing definition of measuring time, when extended to
macroscopic scales, satisfies the Lorentz transformation. Let us consider a
frame S' moving with velocity v with respect to a stationary frame S. We
consider a Planck length at an arbitrary location on the S' frame and at the 
same location on the S frame as is shown in Figure 3. In the S frame, an event 
on the S' frame is measured with relative velocity $c$-v. In
the S' frame, it takes a maximal photon energy pulse $T_{P}^{^{\prime
}}=L_{P}^{^{\prime }}/c$ units of time to pass through the Planck length. However,
this same event, when measured from the same location on the S frame,
measures the time of the energy pulse's passage as $L_{P}^{^{\prime }}/(c-$v%
$)$. Since $L_{P}=L_{P}^{^{\prime }}$ by Postulate 1, and $L_{P}^{^{\prime
}} $ =$T_{P}^{^{\prime }}$ $c$, the time for the maximal photon to traverse $%
L_{P}$ is $T_{P}^{^{\prime }}$ $c/(c-$v$)$. If we let $x=$v$/c$, we obtain
the relation between time observation in the S\ and S' frames at the same
location of Planck length:
\begin{equation}
T_{p}=T_{P}^{^{\prime }}/(1-x)
\end{equation}
\medskip

\begin{center}
\centerline{\includegraphics[bb=56 384 510 688,width=4.5in,
height=3in,keepaspectratio]{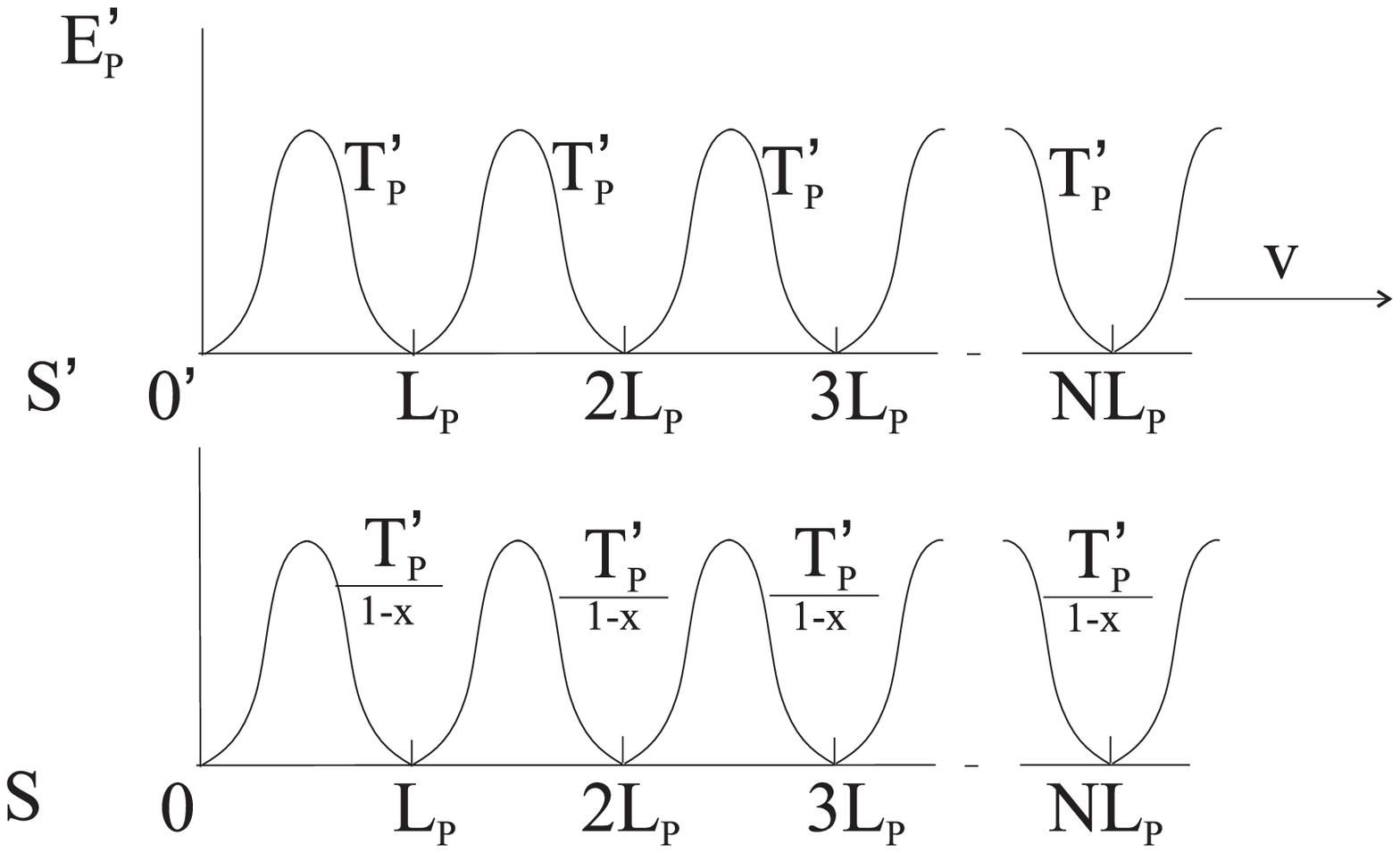}}

\medskip

Figure 3: An event of $N$ energy pulses as viewed in the Moving
Frame S' and in the stationary frame S
\end{center}

\bigskip

We now imagine an event that spans $N$ maximal photons as depicted in Figure
3. Let $iL_{P}$ denote the $i$th Planck length in both frames when the two 
frames are aligned at their respective origins. Although the $i$th 
Planck length is the same in both frames the time duration of a maximal 
photon's passage through the $i$th Planck length is measured according 
to the reference frame. In the 
the S' frame it is $T_{P}^{^{\prime }}$ \ while in S it is $T_{p}$ =$%
T_{P}^{^{\prime }}$ $/(1-x).$ \ How does the observer, stationed at the
origin of S, measure the time that is measured at the $i$th Planck
length location in the S frame? It is reasonable to assume that the energy
associated with $T_{P}^{^{\prime }}$ is reduced and the degree of reduction 
depends on the location on the $x$-axis specified by $i$; the further 
away from the origin, the less energy will be received at the origin,
 and hence less time measured.\

We model this process mathematically. Let $x=$v$/c$  and note that
 the function $1/(1-x)$ in equation (1) is a continuous function 
on $(0,1)$. Let  $ C(0,1)$ denote the space of continuous functions on (0,1).
 Note the function $1/(1-x)$, which defines the time dilation due 
to one maximal photon and the relative motion 
between two frames at the same location. Our objective now is to
 define an
operator $A$, which has the following properties:
1) For any function $f$ in $C(0,1)$, with $f>0$ and $f(0)=1$, the 
iterate $A^{i}f$ denotes
the amount of energy (time) transferred to the origin of S from the 
$i$th Planck 
length location in S. 
2) The total measured time at the origin of S, for large scale times,
 should satisfy the
 Lorentz time dilation transformation. 
That is, if $T'$ is the proper macroscopic time in the $S'$ frame,
 then we want $T= T' \frac{1}{\sqrt{1-x^{2}}}.$ This implies that 
$f^{\ast}(x)= \frac{1}{\sqrt{1-x^{2}}}$ is a fixed point of $A$.
3)We want $f^{\ast}$ to be a stable fixed point of $A$. That is,
 $f^{\ast}$
 should be an attractor in  $ C(0,1)$ so that the approximations 
in using Planck scale lengths are valid.

Let $A:C(0,1)\rightarrow C(0,1)$ denote this
operator and be defined by 
\begin{equation*}
Af(x)=\frac{1+f^{2}(x)\cdot x^{2}}{f(x)}.
\end{equation*}

It is easy to very that $f^{\ast}$ is a fixed point of $A$. It is our 
intention to prove that $A$ satisfies all the desired properties 
and the  iterates of $A$ converge to $f^{\ast}$.

\begin{proposition}
For any function $f(x)>0$ with $f(0)=1$ the iterations $A^{n}f$ converge to
the function 
\begin{equation*}
f^{\ast }(x)=\frac{1}{\sqrt{1-x^{2}}}.
\end{equation*}
\end{proposition}

The proof is a consequence of the following\ three lemmas. Let 
\begin{equation*}
A_{x}(a)=\frac{1+a^{2}x^{2}}{a}\ ,\ \ \ 0<x<1\ ,\ \ \ a>0\ .
\end{equation*}

\begin{lemma}
\label{L:first} a) If $a>\frac{1}{\sqrt{1-x^{2}}}$, then for those $a$
satisfying this inequality we have $A_{x}(a)<a$ and $A_{x}^{2}(a)<a.$

b) If $a<\frac{1}{\sqrt{1-x^{2}}}$, then for those $a$ satisfying this
inequality we have $A_{x}(a)>a$ and $A_{x}^{2}(a)>a.$
\end{lemma}

\begin{proof}
Let $a>\frac{1}{\sqrt{1-x^{2}}}.$ Then, $a^{2}>\frac{1}{{1-x^{2}}}$, or $%
a^{2}-a^{2}x^{2}>1$, and $a>\frac{1+a^{2}x^{2}}{a}.$ To prove the second
statement we continue. We have $a^{2}x^{2}>\left( \frac{1+a^{2}x^{2}}{a}%
\right) ^{2}x^{2}$, or $(1+a^{2}x^{2})\cdot \frac{a}{a}>1+\left( \frac{%
1+a^{2}x^{2}}{a}\right) ^{2}x^{2}$, which means $a>A_{x}^{2}(a)$.

The proof of the statement b) is similar.
\end{proof}

\begin{lemma}
Let $g(x)=\frac{\sqrt{1-x^{2}}}{x^{2}}$. For $x<1/\sqrt{2}$ we have $f^{\ast
}(x)<g(x)$ and for $x>1/\sqrt{2}$ we have $f^{\ast }(x)>g(x)$. For $a$
between the graphs of $f^{\ast }$ and $g$ we have $A_{x}(a)\leq f^{\ast }(x)$
and for the remaining $a$ we have $A_{x}(a)>f^{\ast }(x)$.

In particular: a) If $a<f^{\ast }(x)$ and $x<1/\sqrt{2}$, then $%
A_{x}(a)>f^{\ast }(x)$.

b)If $a>f^{\ast }(x)$ and $x>1/\sqrt{2}$, then $A_{x}(a)>f^{\ast }(x)$.
\end{lemma}

\begin{proof}
This follows by solving the inequality: $A_{x}(a)>\frac{1}{\sqrt{1-x^{2}}}$.
\end{proof}

\begin{lemma}
If $a>0$, then the sequence $\{A_{x}^{n}(a)\}_{n\geq 0}$, $0<x<1,$ converges
to $f^{\ast }(x)=\frac{1}{\sqrt{1-x^{2}}}$.
\end{lemma}

\begin{proof}
First we consider $a>f^{\ast }(x)$.

Let us assume $x<1/\sqrt{2}$. If $a>g(x)$ then the sequence $A_{x}^{n}(a)$
decreases until it goes to or below $g(x)$ at some step $n_{0}$. If $%
A_{x}^{n_{0}}(a)=g(x)$ then the next element $A_{x}^{n_{0}+1}(a)=f^{\ast }(x)
$ and all the following elements have the same value. If $%
A_{x}^{n_{0}}(a)<g(x)$, then the following elements of the sequence
oscillate below and above the value $f^{\ast }(x)$. By Lemma \ref{L:first}
a) the elements above $f^{\ast }(x)$ converge to this value monotonically. The
''below'' elements of the sequence also converge to the same limit since $%
A_{x}$ is continuous.

For $x>1/\sqrt{2}$ the sequence $A_{x}^{n}(a)$ is decreasing and converges
to $f^{\ast }(x)$ monotonically.

Now, let $a<f^{\ast }(x)$. If $x<1/\sqrt{2}$, then $A_{x}(a)>f^{\ast }(x)$
and we have convergence by the first part of the proof.

Let $x>1/\sqrt{2}$. If $a\leq g(x)$, then again $A_{x}(a)>f^{\ast }(x)$ and
we have convergence by the first part of the proof. If $a>g(x)$, then the
sequence $A_{x}^{n}(a)$ is increasing and converges to $f^{\ast }(x)$
monotonically.
\end{proof}

The following theorem is the consequence of Proposition 1.

\begin{theorem}
Let $f(x)=\frac{1}{1-x}$. We have the convergence of the averages 
\begin{equation*}
\frac{1}{N}\left( f+A(f)+A^{2}(f)+\dots +A^{N-1}(f)\right) \rightarrow
f^{\ast }.
\end{equation*}
\end{theorem}

This means that 
\begin{equation*}
\frac{T_{P}^{^{\prime }}}{N}[1/(1-x)+A^{1}(1/(1-x)+A^{2}(1/(1-x)+\dots
+A^{N-1}(1/(1-x)]\rightarrow \frac{T_{P}^{^{\prime }}}{\sqrt{1-x^{2}}}
\end{equation*}\
or
\begin{equation*}
T_{P}^{^{\prime }}[1/(1-x)+A^{1}(1/(1-x)+A^{2}(1/(1-x)+\dots
+A^{N-1}(1/(1-x)]\rightarrow \frac{N T_{P}^{^{\prime }}}{\sqrt{1-x^{2}}}
\end{equation*}\

that is, the total time observed at the origin of S approaches $\frac{%
NT_{P}^{^{\prime }}}{\sqrt{1-x^{2}}}$ as the number of maximal photon pulses 
increases to $\infty.$ But $NT_{P}^{^{\prime }}$ is the proper time 
observed in the S' frame. 
Hence we have derived the Lorentz
transformation for time dilation.
\medskip

It is of interest to know how the measured dilation times approaches the 
Lorentz transformation. The following lemma shows that at all scales 
the dilation function stays above the graph of $\frac 1{\sqrt{1-x^{2}}}$.
\medskip

\begin{proposition}
If $f\ge f^*$, then the sequence of averages stays above the limit function $%
f^*$: 
\begin{equation*}
\frac 1 N \sum_{n=0}^{N-1}A^n(f)\ge f^*\ .
\end{equation*}
\end{proposition}

\begin{proof}
As we showed in Lemma 2, for $x>1/\sqrt{2}$ and $a>f^*(x)$ the
 sequence $A_x^n(a)$ decreases monotonically and is above
$f^*(x)$. The statement of the Lemma follows.

For $x<1/\sqrt{2}$ and $a>f^*(x)$ the sequence $A_x^n(a)$ decreases 
monotonically for some time and then starts to oscillate below and
 above $f^*(x)$. We will prove that 
$$ a > f^*(x) \text{  and  } A_x(a)< f^*(x) \ \ \Longrightarrow
 \ \  f^*(x)- A_x(a)<a-f^*(x)\ .$$
This implies the statement of the Lemma.

We want to show $$\frac 1 {\sqrt{1-x^2}} - \frac {1+a^2x^2}{a}
< a-\frac 1 {\sqrt{1-x^2}},$$
or $$ a^2(1+x^2)-a\frac 2 {\sqrt{1-x^2}}+1>0.$$
Standard calculations show that this holds for $a>\frac 1 {\sqrt{1-x^2}}$.
\end{proof}

\medskip
\begin{center}
\centerline{\includegraphics[bb=0 0 400 400,width=4.17in,
height=4.17in,keepaspectratio]{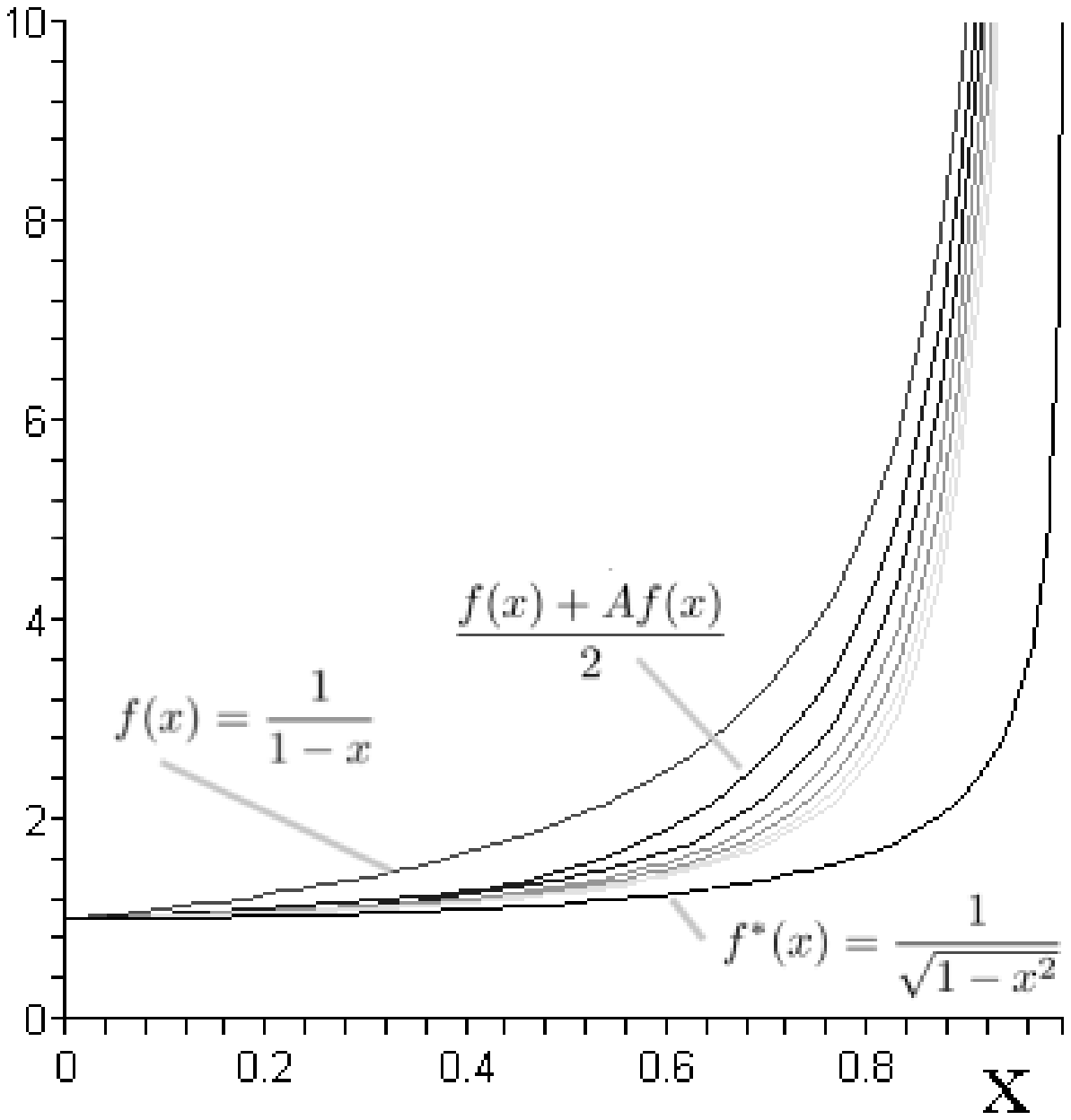}}

Figure 4: Averages $\frac 1 N \sum_{n=0}^{N-1}A^n(f)$, 
$N=1,2,\dots,7$, for $f(x)=\frac 1{1-x}$(top curve) and the
limit function $f^*(x)=\frac 1{\sqrt{1-x^2}}$ (bottom curve).

\end{center}

\medskip

Notes:
1.) Although in theory it is possible for maximal photons to exist, one may 
ask why they have not been observed. We suggest three possible answers:
 a)we do not have the experimental tools that can measure such small 
wavelengths, b) if the Planck-Einstein equation applies at the Planck 
length scale, then we do observe $E_{P}$ in the form of time that we 
can measure, and c) on a more philosophical note, the energy $E_{P}$ 
is the energy that allows a Planck length of space to be observed in time.

2.)  Is time a dimension? The approach of this note argues against this. Time is not regarded as a coordinate. From our perspective, time is no different than color. Neither time nor color are independent variables; they are attributes of space. During an event, time accrues to Planck lengths in the form of energy. Thus, time is merely a number at every spatial location. When considered over many events, time at a Planck length location may be a fractal, due to the gaps between events and the various scales incurred by the duration of events.

3.) The cosmic ray paradox is concerned with why we are able to observe UHE particles. The tenor of this note is that such particles are ubiquitous and observable in the form of measurable time. But not all such particles adhere to Planck lengths of space and it is these particles that engender the cosmic ray paradox. We may ask why we cannot observe more of these UHE particles that are transformed into time. These particles create the observer's awareness of time and hence his consciousness. This process is not measurable directly by the observer much as the high energy jolt of a defibrillator cannot be experienced by the person who is unconscious. It is only the much lower energies associated with living that can be experienced afterward.

4.) Dark energy may be energy carried by time much as ordinary energy is carried by mass.

\end{document}